# Experimental Realization of the 1D Random Field Ising Model

N. S. Bingham[1], S. Rooke[2], J. Park[2], A. Simon[1], W. Zhu[1], X. Zhang[1,2], J. Batley[3], J. D. Watts[3,4], C. Leighton[3], K. A. Dahmen[2], P. Schiffer[1,2,5]

1. Department of Applied Physics, Yale University, New Haven, CT 06511 USA

2. Department of Physics, University of Illinois at Urbana-Champaign, Urbana, IL 61801 USA

3. Department of Chemical Engineering and Materials Science, University of Minnesota, Minneapolis, Minnesota 55455, USA

4. School of Physics and Astronomy, University of Minnesota, Minneapolis, Minnesota 55455, USA

5. Department of Physics, Yale University, New Haven, CT 06511 USA

**Abstract**

We have measured magnetic-field-induced avalanches in a square artificial spin ice array of interacting nanomagnets. Starting from the ground state ordered configuration, we imaged the individual nanomagnet moments after each successive application of an incrementally increasing field. The statistics of the evolution of the moment configuration show good agreement with the canonical one-dimensional random field Ising model. We extract information about the microscopic structure of the arrays from our macroscopic measurements of their collective behavior, demonstrating a process that could be applied to other systems exhibiting avalanches.

Avalanche responses to a slow driving field are ubiquitous in nature, appearing in systems as diverse as magnetic domains, superconducting vortices, earthquakes, landslides, power grids, and the stock market [1,2,3,4,5]. The physics of these systems is particularly complex, due to the presence of disorder, the intrinsically metastable nature of the phenomena, and the proximity to instabilities.

An important theoretical model for understanding avalanches in their simplest form is the one-dimensional random field Ising model (1D-RFIM) [6,7], which is broadly applicable to many systems [8,9,10,11]. A simple experimental realization of the 1D-RFIM has, however, been elusive. Typical experimental systems are impacted by complications such as long-range forces, higher dimensional character, or inaccessibility of the critical regime where fluctuations are large. Furthermore, very few systems are amenable to experimental imaging of full avalanches with sufficient resolution to precisely characterize their scale.

We report a study of avalanche behavior in arrays of coupled single-domain ferromagnetic nanoislands. These arrays, known as artificial spin ice (ASI), have the unusual property of being designable at the microscopic scale and also accessible to high-resolution imaging at the level of the individual nanomagnet moments. Such systems display a wide range of interesting behavior, including unusual ground states and magnetic-monopole-like excitations [12,13]. Avalanche-like phenomena have previously been examined in ASI arrays through studies of reversal behavior from one polarized state to another [14,15,16,17,18,19,20,21,22,23,24,25]. These included avalanche statistics and the relation of avalanche phenomena to the formation of monopole-antimonopole pairs. The avalanches in these studies had significant two-dimensional character, associated with the coupling of the moments transverse to the field direction and starting from a fully polarized state.

We have studied linear avalanche-like reversal in the polarization of the nanomagnet moments in an appropriately oriented ASI array. By analyzing the statistics of the reversed moments, we find that this system provides a clear experimental manifestation of the 1D-RFIM. Based on the capacity to image individual moment orientations after successive avalanches, we also demonstrate a generally applicable method to extract the underlying random field distribution. Significant additional data, derivations, experimental techniques and references [26] are included in the Supplemental Information.

We examined avalanche phenomena in an ASI system designed to produce strictly one-dimensional avalanches. Specifically, we studied a rotated version of the canonical square ice system (see figure 1a [27]), where each island was a single ferromagnetic domain with magnetization oriented along the long axis by shape anisotropy. Because of the rotation of our structure, the islands formed vertical columns (figures 1 d,e), with the island axes, and thus the magnetic moments, oriented at 45° from the column direction (similar to the geometry in [15,28]). Our square arrays had size $L$ = 70, 80, and 100, where $L$ is the number of islands on each side of an array. The permalloy, $Ni_{80}Fe_{20}$, islands were patterned by e-beam lithography with a nominal island size of 220±11 nm x 80±8 nm with nominal thickness of 15 nm, a lattice constant of $a$ = 320 nm (defined in figure 1d). The islands were capped with 3 nm of Al (forming $Al/AlO_x$) to prevent oxidation. To demonstrate reproducibility, we measured four arrays with $L$ = 70, and three arrays each for $L$ = 80 and 100, and the results are qualitatively consistent across all arrays. For the purpose of indicating which data are from which array, the arrays are labelled A,B,C,D, for $L$ = 70, and A,B,C, for $L$ = 80 and 100.

Our arrays, as-grown, had all their moments arranged in the ordered ground state of the square ice system, a phenomenon observed previously as the formation of extended domains in larger arrays [29]. In the ground state, alternating columns of nanomagnets have opposite polarizations, either up or down in figure 1 (tilted at 45 degrees due to the island rotation). Application of a field along the columns only flips moments that are anti-aligned with the field, i.e., those in every other column. This arrangement enforces the one-dimensional nature of the magnetic reversal. The measured magnetization as a function of applied field for a very large array ($L$ = 10,000), is shown in figure 1c. The steep part of the curve is associated with moments flipping to align with the external field.

Our choice of island thickness ensured that the moments were thermally stable and could be measured by magnetic force microscopy (MFM) without alteration of the moment state (typical data are shown in figure 1b). We took data by applying a magnetic field ($H_{ext}$), reducing the field to zero, and then mapping the orientation of the moments with MFM. We repeated this process for gradually increasing values of $H_{ext}$ until the system was fully polarized, allowing us to track successive increases in the number of flipped moments. The field was aligned with the array using Au alignment bars and a precision rotation stage with ± 0.5° resolution (SQUID magnetization measurement indicated that angles up to 10° did not impact the data). We indicate the studied values of $H_{ext}$ on figure 1c, and additional data and our digitization technique are described in the Supplementary Information (section SI-2).

In figure 2, we show the measured moment configurations from one of the $L$ = 70 arrays for a set of successively increasing values of $H_{ext}$. The flipped moments are indicated in red, and a cursory examination reveals that they occurred in clusters, growing in avalanche-like steps with each increasing value of $H_{ext}$. The size of a cluster of flipped moments, $S$, can be defined as the total number of consecutive flipped moments along a column at a given $H_{ext}$. Because we imaged in zero field at remanence, we did not examine the dynamics of the moment flipping process (which is expected to occur many orders of magnitude faster than MFM imaging) [15].

The MFM maps of moments for each value of $H_{ext}$ were converted to a distribution of cluster sizes $D(S,H_{ext})$, which is plotted below in figure 5. In figure 3, we show an average of the distributions of cluster sizes, $D_{avg}(S)$ for all of our arrays for each of our three different array sizes with the data binned logarithmically as a function of $S$, for $S < L$. Note that $D_{avg}(S)$ is averaged over all values of $H_{ext}$ for which island moments reversed, and over samples with the same $L$, then normalized by the number of moments initially aligned antiparallel to $H_{ext}$, i.e., $L^2/2$. For $D_{avg}(S)$, we also discarded any system-spanning clusters of size $S = L$. The collapse of the data in figure 3 indicates the presence of a robust and broad range of length scales, and that clusters smaller than $L$ are not significantly affected by system size effects. The observed functional form is consistent with our theoretical predictions, as discussed below.

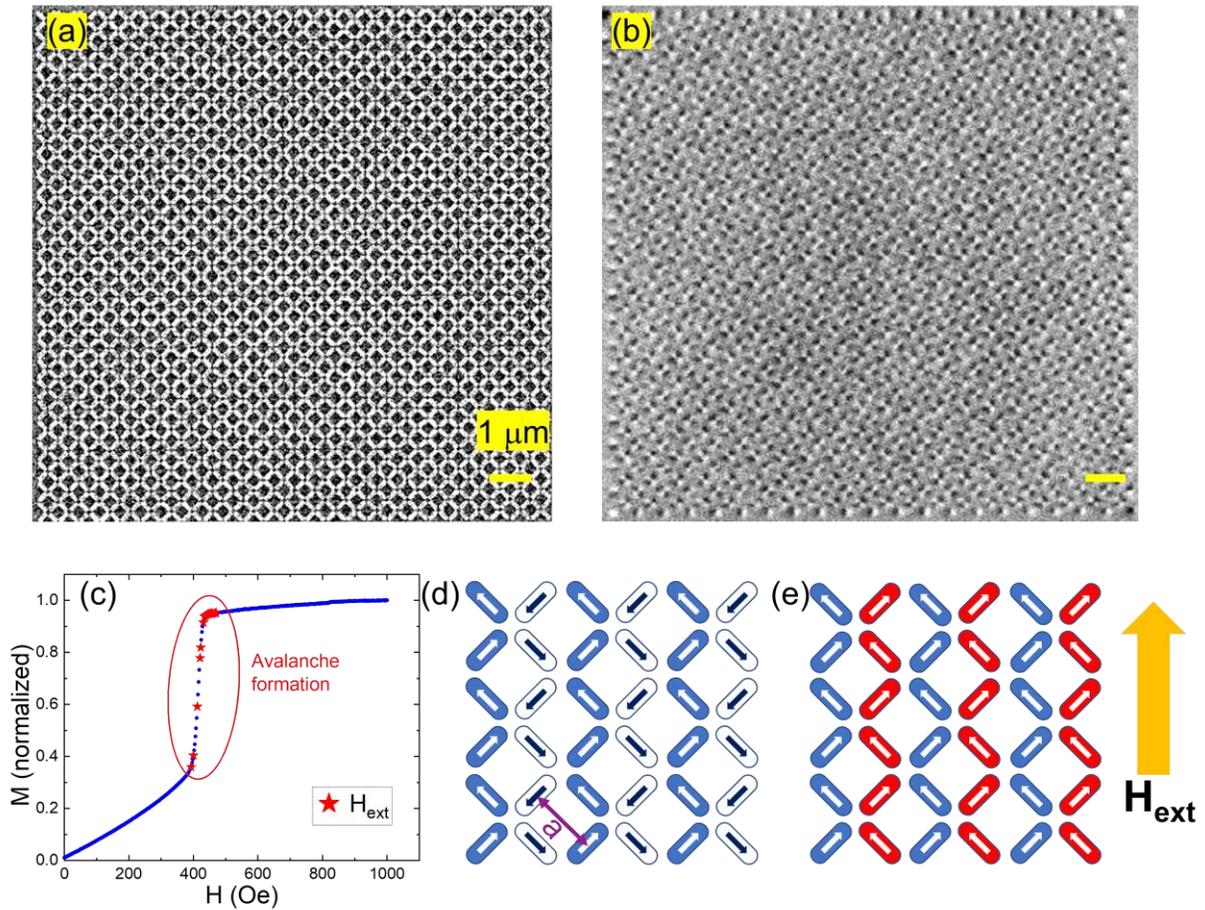

**Figure 1. (a)** Atomic force microscope image of the rotated square lattice. **(b)** Magnetic force microscope (MFM) image of the ground state of $L = 70$ array, Sample C. **(c)** Normalized magnetization measured with a Quantum Design SQUID magnetometer (MPMS 3) as a function of applied field for an $L = 10,000$ array (blue circles), and the fields (red stars) corresponding to the values of $H_{ext}$ used in the MFM studies. **(d)** Schematic of the lattice with arrows indicating initial moment configuration, and the yellow arrow indicating the direction of $H_{ext}$. The lattice constant for the structure is also indicated as *a*. **(e)** Schematic of the lattice with arrows indicating the polarized state moment configuration.

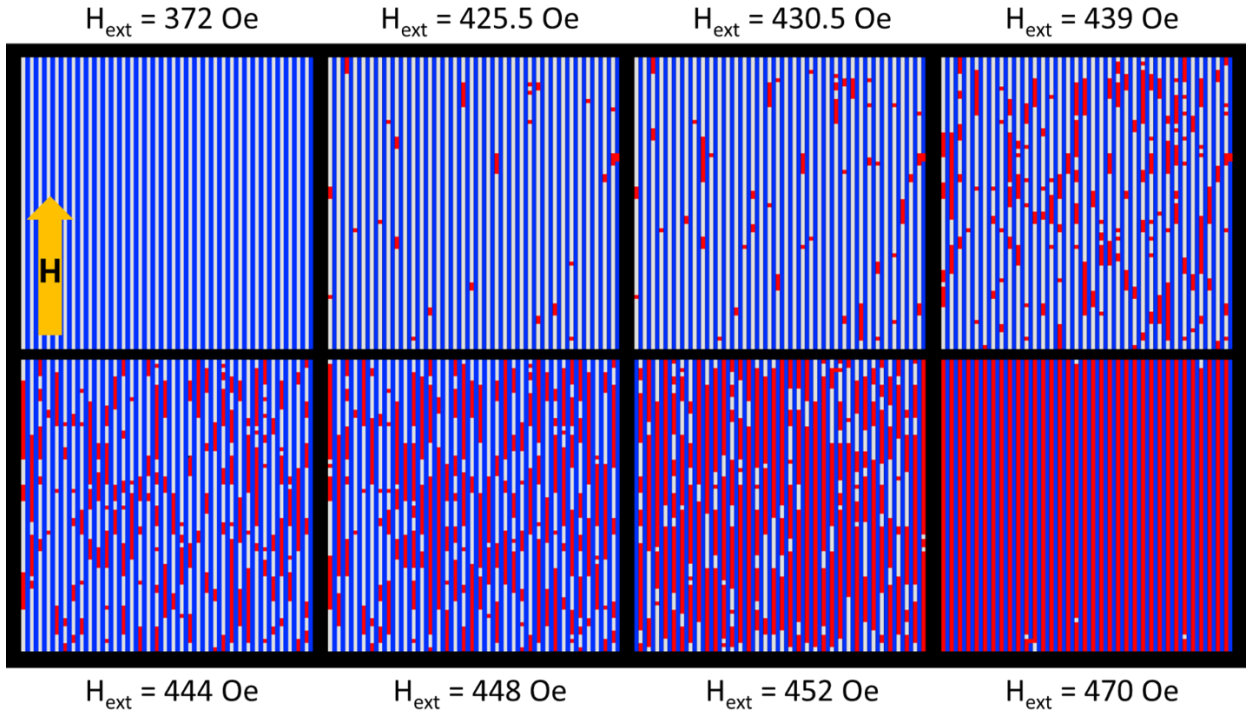

**Figure 2.** Digitized moment maps for the field evolution of avalanche growth for an *L* = 70 array, Sample C. The blue stripes represent columns of moments that were aligned with the $H_{ext}$ direction in the initial state of the system. The white points represent moments that were anti-aligned with $H_{ext}$ in the initial configuration. The red points represent moments that flipped to align with the field.

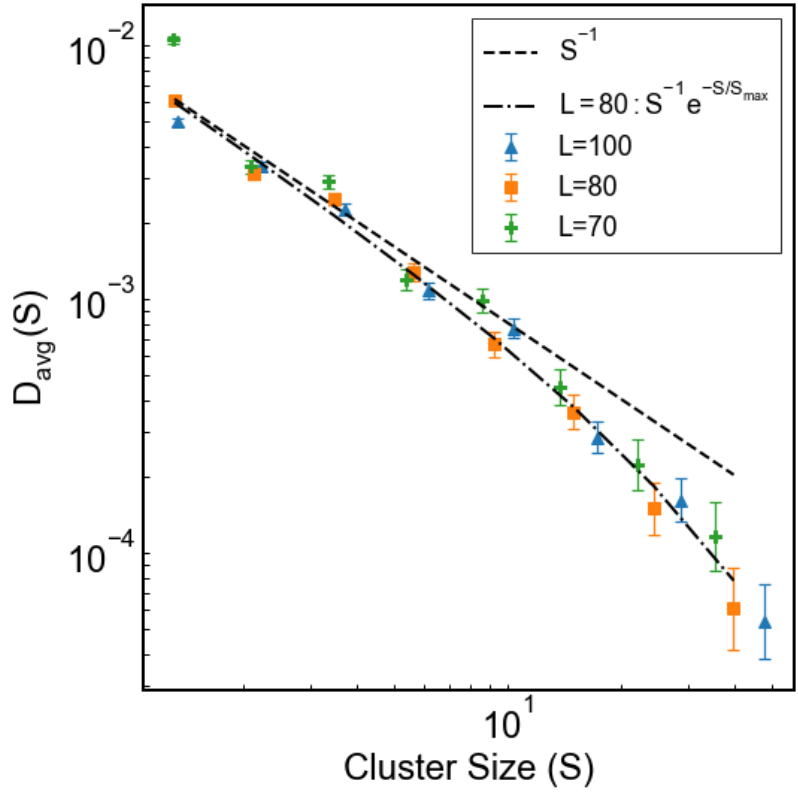

**Figure 3:** Averaged distribution, $D_{avg}(S)$ for clusters smaller than the system size ($S < L$), averaging over all samples with a given $L$. In an infinite system this quantity is expected to follow a $S^{-1}$ power law (dashed line) as derived in the Supplementary Information (section SI-7). In a finite system it is modified by a fitted exponential cutoff (Supplementary Information, section SI-5). A fit to this form for $L = 80$ is shown (dash-dot line).

The islands in ASI have disorder associated with the polycrystalline nature of the permalloy and with imperfections in the lithography (see Supplementary Information, section SI-1). The resulting disorder should be reflected in the energetics associated with the nucleation of moment reversals as well as the interactions among neighboring moments, and it has been shown previously to impact the physics of ASI [30,31,32,33]. Because of the effective one-dimensional nature of our structure, the 1D-RFIM is a natural model for our experimental data that can reflect the underlying physics through a simple parameterization of the disorder.

The 1D-RFIM for Ising-like spins has the general Hamiltonian [6,7]:

$$\mathcal{H} = -\sum_{i,j} J_{ij}\, s_i s_j - \sum_i (H_{ext} + H_i) s_i. \qquad (Eq.\,1)$$

Here, $s_i$ denotes a particular spin, $J_{ij}$ the dipolar interaction strength between spins *i* and *j*, $H_{ext}$ denotes the external field, and $H_i$ denotes an effective intrinsic field that varies randomly between sites (see Supplementary Information, section SI-1). A simple Ising spin model where each spin can only take discrete values of ± 1 is applicable because of the strong shape anisotropy of the islands [6,7,27,34,35].

We take $J_{ij} \equiv J > 0$ for the nearest neighbors along a column of moments that can be flipped by $H_{ext}$ (the white and red moments in figures 1d,e respectively). The nearest neighbor moments in adjacent columns (the blue moments in those figures) were aligned with $H_{ext}$ and remained fixed during an avalanche. Thus, the interaction field associated with them can effectively be viewed as a small shift in the mean of $H_i$ (longer-range interactions among more distant columns appear to be negligible, as discussed in the Supplementary Information, section SI-3). We can then view the Hamiltonian as:

$$\mathcal{H} = -\sum_{<i,j>} \left[ s_j J + H_{ext} + H_i \right] s_i. \qquad (Eq.\,2)$$

Based on the central limit theorem, the random field strengths are typically assumed to follow a Gaussian distribution [6,35]:

$$\phi(H_i) = \frac{1}{\sqrt{2\pi R^2}} e^{-(H_i - h)^2 / 2R^2} \qquad (Eq.\,3)$$

where the width of the distribution, $R$, represents disorder strength, and $h$ is the center of the distribution. We use the Gaussian distribution in the discussion below, and we verified that it represents the data better than a Lorentzian distribution (see Supplementary Information, section SI-11). We can include all randomness in $H_i$, because, in one dimension, any randomness in $J$ can be mapped onto an additional randomness in $H_i$ for homogeneous nearest neighbor coupling [35].

For our experimental configuration, the moments that were flipped by the field were initially oriented opposite $H_{ext}$ (downward in figure 1). As $H_{ext}$ was increased beyond the intrinsic field and the interaction term, moments associated with weaker random fields flipped their orientations. This impacted the energetics for nearby moments along the same column, and thus multiple moments typically flipped together in an avalanche for a particular value of $H_{ext}$, until the avalanche hit a particularly strong random field, collided with a separate cluster of already-flipped moments in the same column, or reached the sample edge. The probability, $p(H_{ext})$, that a growing cluster that has propagated to site [$i$] continues to a site [$i$ + 1] depends on $H_{ext}$ through the relation:

$$p(H_{ext}) = P(H_{i+1} > -H_{ext} - \alpha J), \qquad (Eq.\,4)$$

where we approximate the interaction term as $\alpha J$, with the exact value of α depending on geometric factors that do not affect our analysis. Note that, at the edge of a propagating avalanche, the interaction term is constant, as there is always a previously flipped moment behind the edge of the avalanche as it propagates, and a moment that has yet to flip ahead of the propagating avalanche.

To compare with experiment, we obtain $p(H_{ext})$ from the measured distribution of cluster sizes, $D(S,H_{ext})$. We neglect collisions of the avalanches with other clusters or the sample edge, a reasonable approximation for sufficiently low $H_{ext}$ (discussed below). A cluster always consists of at least one moment-flip and has $S = 1$ with probability $\propto (1-p)^2$, which is the probability that both nearest neighbors of the single flipped moment remain unchanged. Likewise, as $p$ is constant at the edge of a growing cluster of flipped moments, a cluster is of exactly size $S$ with probability $Ap^{S-1}(1-p)^2$, where $A$ is a normalization constant, defined as $A = (1-p)^{-1}$. As a result, the cluster size distribution is given by:

$$D(S, H_{ext}) = (1-p)p^{S-1}. \qquad (Eq.5)$$

We analyze the cluster data through the complementary cumulative distribution function (CCDF), which is the distribution of clusters of size $S$ or greater for a given $H_{ext}$. From Eq. 5 we obtain the following CCDF:

$$C(S, H_{ext}) = [p(H_{ext})]^{S-1}. \qquad (Eq.6)$$

We fit Eq. 6 to our experimental data at each value of $H_{ext}$ (example fits shown in Figure 4a) to find individual values of $p(H_{ext})$ (Figure 4b). We find good agreement with the model, especially for small values of $p$, where finite size effects are minimized.

If the random field is normally distributed, we expect:

$$p(H_{ext}) = P(H_{i+1} > -H_{ext} - \alpha J) = \int_{-\infty}^{H_{ext}} \frac{1}{\sqrt{2\pi R^2}} e^{\frac{-(-x-h')^2}{2R^2}} dx \qquad (Eq.7)$$

where, $h' = h + \alpha J$. From the values for $p(H_{ext})$, we can obtain the parameters $h'$ and $R$ in the distribution of random fields (fit in figure 4b, and additional fits in Supplementary Information, section SI-8). Reconstructions of the distributions of random fields, using the values $h'$ and $R$ extracted from the same fitting process of $C(S,H_{ext})$ and $p(H_{ext})$ for the nominally identical $L = 70$ arrays, are shown in figure 4c. The distributions show small variations as might be expected from, e.g., imperfections in the lithography.

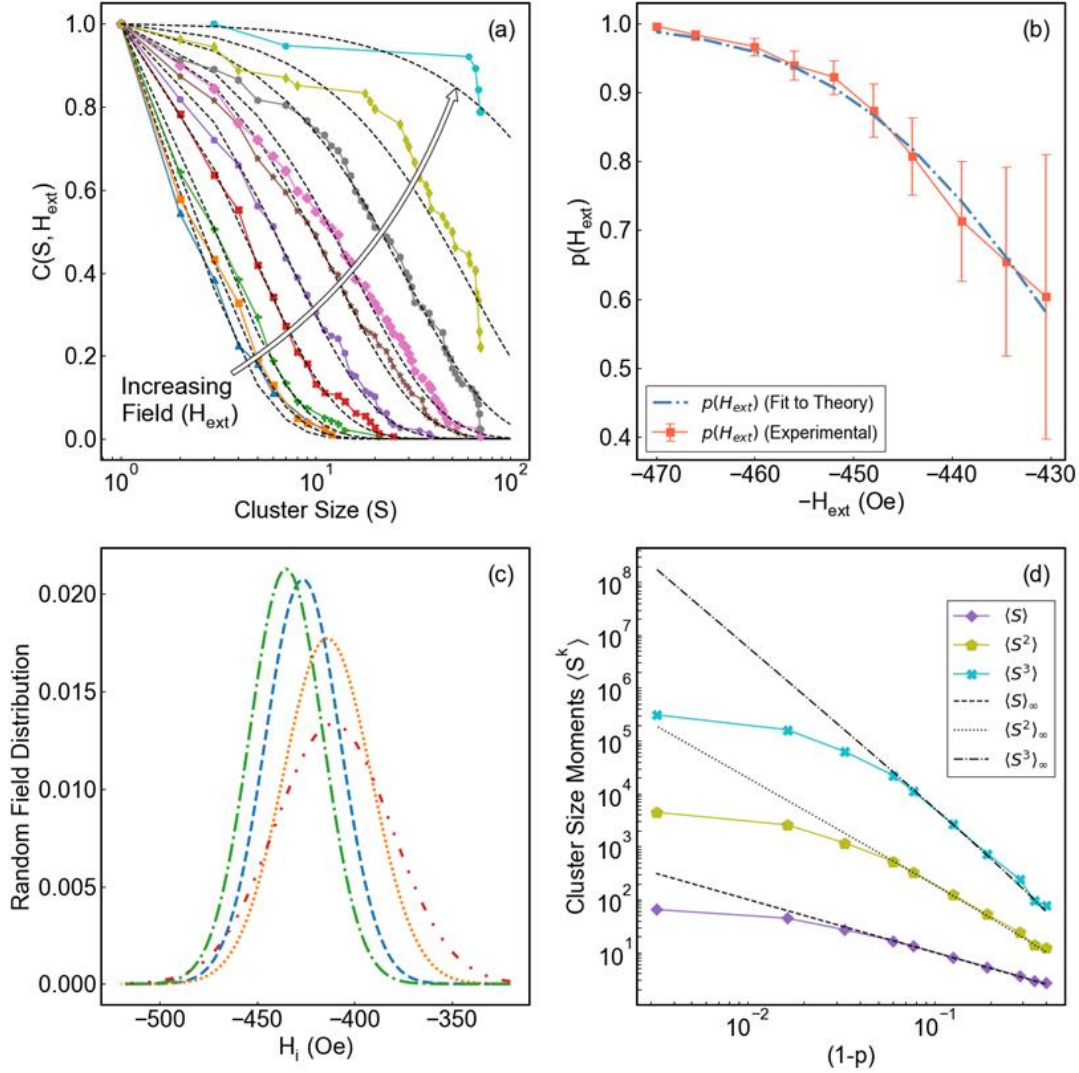

**Figure 4: (a)** Complementary Cumulative Distribution Functions (CCDFs) of cluster sizes for a single $L = 70$ array, (Sample A), and $H_{ext}$ = 430.5, 434.5, 439, 444, 448, 452, 456, 460, 466, and 470 Oe. The dotted lines are fitted model predictions. **(b)** The experimentally recovered values of $p(H_{ext})$ calculated from the data and fits in (a). The blue line is predicted from Eq. 7 with fitted values of $R = 19.2 \pm 1.0$ and $h' = 426.6 \pm 0.7$ Oe. **(c)** Gaussian distributions of random fields (Eq. 3) using the fitted values of $R$ and $h'$ for all four arrays with $L = 70$ (Samples A, B, C, and D). **(d)** Moments of the cluster size distribution plotted as a function of 1-$p$, for the single L = 70, Sample A array, where dashed lines show the model prediction as discussed in the text. Here, we include fields in which spanning clusters are present, so that the deviation from the model due to finite size at smaller values of 1-$p$ is apparent.

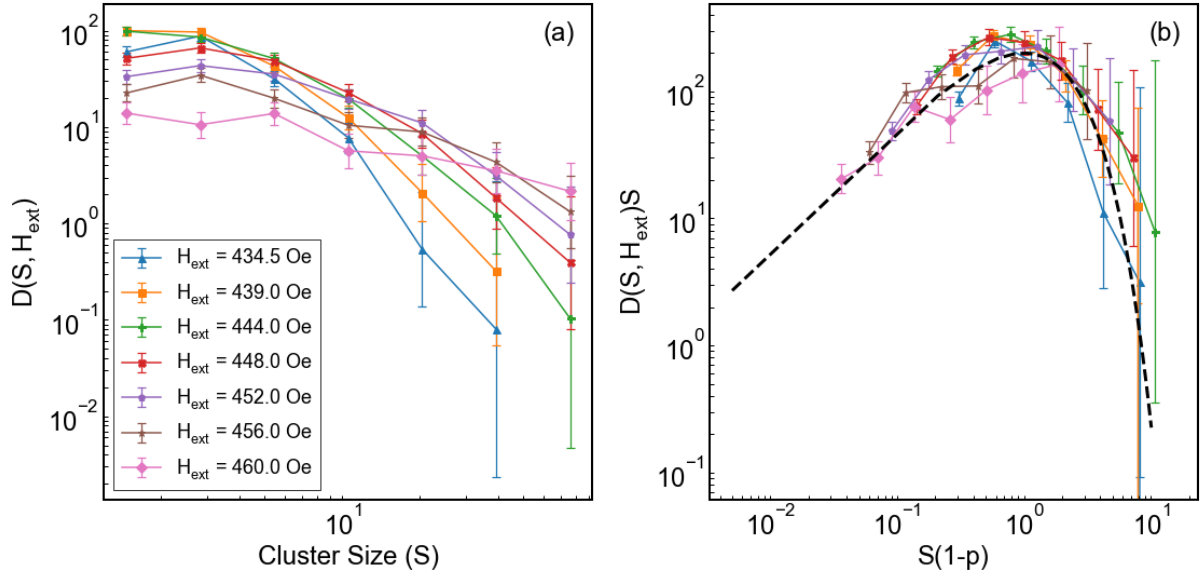

**Figure 5: (a)** Distribution of cluster sizes measured for various values of $H_{ext}$, averaging data over all three arrays with $L = 100$ (A, B, and C). Here, we plot this distribution using values of $H_{ext}$ for which there is substantial island moment flipping, but do not plot the distribution for fields in which spanning clusters dominate. **(b)** The same data after scaling the axes compared to the predicted scaling function $Axe^{-x}$ in black, as described in the text.

Using the fits to $p(H_{ext})$, we can additionally test some of the assumptions used in our derivations. In figure 4d, we plot moments of the cluster size distribution, ($\langle S^n \rangle$) for one array. As shown in the Supplementary Information (section SI-6), the $n^{th}$ moment on an infinite lattice is expected to be given by:

$$\langle S^n \rangle \equiv \sum_{S=1}^{\infty} S^n D(S, H_{ext}) = (1-p)[(\frac{d}{dp})p]^n \frac{1}{1-p}.$$

Here, we consider all external fields for which moments flipped. The data show good agreement with this expression for $p \rightarrow 0$, indicating that the nearest neighbor cluster description is justified. Conversely, the lack of agreement for $p \sim 1$ is expected, since that regime is near full polarization, where finite-size effects and cluster collisions become non-negligible.

In figure 5a, we plot the distribution $D(S,H_{ext})$, without averaging over $H_{ext}$. In figure 5b, we scale these data, using our fitted values of $p(H_{ext})$, with the approximate form $D(S,H_{ext}) \sim S^{-1}F[S(1-p)]$ where $F[S(1-p)] \sim S(1-p)e^{-S(1-p)}$, as derived for an infinite system near $p \approx 1$ in the Supplementary Information (section SI-7). As can be seen in the figure, the data collapse is good and follows the predicted scaling form, indicating that our model is valid for most values of $H_{ext}$, even in the presence of finite size effects and cluster collisions. Thus, the random field in our analysis dominates corrections from these effects, except at the largest $H_{ext}$, where all moments are polarized. We also note that, on an infinite lattice, we expect $D_{avg}(S)$ (plotted in Figure 3) to follow an $S^{-1}$ power law (see Supplementary Information, section SI-7). We fit the data in Figure 3 to this form with an exponential prefactor, i.e., $D_{avg}(S) \sim e^{-S/S_{max}}S^{-1}$, where the cutoff $S_{max}$ is a fitting parameter that reflects a finite correlation length for the clusters in a finite system, further validating our approach.

The above results are qualitatively equivalent across multiple samples, as demonstrated in several sections of the Supplementary Information. We therefore conclude that the square ASI system provides a robust experimental realization of the non-equilibrium 1D-RFIM. While the RFIM is a seminal model in statistical mechanics, experimental extractions of distributions of disorder strengths (such as the underlying random fields or, equivalently, random anisotropies) have been quite rare. The above results demonstrate a general method for reconstructing the underlying distribution of random fields from the distribution of cluster sizes. The results of this reconstruction, obtained from one specific history of the sample, allows us, in principle, to predict the system's behavior under all possible histories. This possibility is supported by previous findings of return point memory in square ASI [36].

Although we are primarily concerned with the 1D-RFIM, our methods should be easily transferable to spin lattices of higher dimension. They should be especially pertinent to quasi-one-dimensional and two-dimensional systems, where similar imaging techniques can be utilized. Furthermore, by careful control of the ASI fabrication, we should be able to probe the statistical properties of the 1D-RFIM by controlling the sources of disorder. Importantly, our methodology for measuring the successive states of growing clusters should be accessible in many other systems where avalanche-type behavior is observed and can be measured, e.g., materials testing and hazard prevention. The resulting applicability of the well-studied 1D-RFIM to understanding of such behavior should open the door to a range of both fundamental and applied studies.

**AUTHOR CONTRIBUTIONS AND FUNDING ACKNOWLEDGEMENT**

Lithography, experimental measurements, and data analysis were conducted by NB, JP, AS, WZ and PS at Yale University and the University of Illinois at Urbana-Champaign, funded by the US Department of Energy, Office of Basic Energy Sciences, Materials Sciences and Engineering Division under Grant No. DE-SC0010778 and Grant No. DE-SC0020162. Sample growth was performed by JB, JW, and CL at the University of Minnesota and was supported by NSF through Grant No. DMR-1807124 and DMR-2103711. Theoretical framing of the issues, as well as derivations, calculations, and additional data analysis were conducted by SR and KD at the University of Illinois at Urbana-Champaign. The authors are grateful to Cristiano Nisoli, Vincent Crespi, and Paul Lammert for helpful discussions.